\documentclass[aps,prl,twocolumn,showpacs,preprintnumbers,a4paper]{revtex4} 

\usepackage{graphics}
\usepackage{epsf}
\usepackage{eufrak}
\usepackage{epsfig}

\begin{document} 


\title{Universal behavior of crossover scaling functions for 
continuous phase transitions}

\author{S. L\"ubeck}
\affiliation{Weizmann Institute of Science, 
Department of Physics of Complex Systems, 
76100 Rehovot, Israel,\\
Institut f\"ur Theoretische Physik,
Universit\"at Duisburg-Essen, 
47048 Duisburg, Germany}

\date{Received 3\,March, 2003, published 27\,May 2003}

\begin{abstract}
We consider two different systems exhibiting 
a continuous phase transition into an absorbing 
state.
Both models belong to the same universality class, i.e.,
they are characterized by the same scaling
functions and the same critical exponents.
Varying the range of interactions we
examine the crossover from the mean-field-like
to the non-mean-field scaling behavior.
A phenomenological scaling form is applied in order
to describe the full crossover region which spans
several decades.
Our results strongly supports the hypotheses
that the crossover function is universal.
\end{abstract}



\preprint{{\it Physical Review Letters} {\bf 90}, 210601 (2003)}

\maketitle


The critical behavior of a system exhibiting a second
order phase transition with non-mean-like scaling 
behavior is strongly affected by the
range of interactions.
The longer the range of interactions the stronger
will be the critical fluctuations reduced.
In the limit of infinite interactions the system 
is characterized by the mean-field scaling behavior.
But according to the well known Ginzburg 
criterion~\cite{GINZBURG_1}, 
mean-field-like behavior occurs even
for finite interaction ranges sufficiently far 
away from the critical point.
A crossover to the non-mean-field scaling
behavior takes place if one approaches the transition
point.
Although crossover phenomena are well understand 
in terms of competing fix points of the corresponding
renormalization group approaches (see for instance~\cite{YEOMANS_1}),
some aspects of crossover phenomena
are still open.
For instance it is an open question whether the
so-called effective exponents fulfill 
certain scaling relations over the entire crossover 
region (see~\cite{CHANG_1,LUIJTEN_1,LUIJTEN_2,MARQUES_1} and 
references therein).
A second open question is the main theme of this
paper and addresses the universality of the crossover 
scaling functions.
The range where the universal critical scaling
behavior applies is usually restricted to a small vicinity 
around the critical point.
Therefore it is questioned that the full crossover
region, that spans several decades 
in temperature or conjugated field, 
can be described in terms of universal scaling
functions.
Renormalization group approaches 
predicted a non-universal behavior if one
uses finite cutoff lengths whereas 
infinite cutoff lengths (which corresponds to an
unphysical vanishing molecular size) lead to a universal 
crossover behavior (see for instance~\cite{ANISIMOV_1,BELYAKOV_1}).
On the other hand the experimental situation is also 
unclear since measurements over the whole
crossover region are difficult and accurate
results are rare (see~\cite{LUIJTEN_1} for a
short discussion).
Thus several attempts were performed in order
to address this question via numerical simulations.
For instance the two- and three-dimensional Ising 
model with various interaction ranges is considered
in a series of papers~\cite{LUIJTEN_1,LUIJTEN_2,MON_1,LUIJTEN_3}.
Using a sophisticated cluster algorithm for long-range
interactions it was possible to cover the full 
crossover region.
In particular a collapse of the susceptibility for
different values of the interaction range was observed.
Thus the crossover can be described by a single scaling function
in agreement with renormalization group approaches.
But this result does not present an evidence that this scaling
function is universal, since only one system of a given
universality class was considered.

The purpose of this paper is to demonstrate via numerical
simulations that the crossover from non-mean-field
to mean-field-like scaling behavior can be represented by 
universal functions.
We therefore consider two different
systems exhibiting a continuous phase transition;
both belong to the same universality class.
The dynamics of the models is characterized by
simple particle hopping processes, i.e., 
various interaction ranges can be easily implemented and 
highly accurate data are available.
In this way it is possible to observe the full
crossover region. 
Notice that we focus in our investigations on the 
particular universality class of absorbing phase transitions 
only for technical reasons.
The demonstrated universality of crossover
scaling functions can be applied to continuous phase
transitions in general.


The first considered model is the so-called conserved 
lattice gas (CLG) which
was introduced in~\cite{ROSSI_1}.
In the CLG lattice sites may be empty or occupied
by one particle.
In order to mimic a repulsive interaction a given particle
is considered as active if at least one of its
neighboring sites on the lattice is occupied by another
particle.
If all neighboring sites are empty the particle remains
inactive.
Active particles are moved in the next update
step to one of their empty nearest neighbor sites,
selected at random.

The second model is the so-called conserved transfer
threshold process (CTTP)~\cite{ROSSI_1}.
Here, lattice sites may be empty, occupied by one particle,
or occupied by two particles.
Empty and single occupied sites are considered as
inactive whereas double occupied lattice sites are 
considered as active.
In the latter case one tries to transfer both particles
of a given active site to randomly chosen empty or single
occupied nearest neighbor sites.

In our simulations (see~\cite{LUEB_22,LUEB_24} for 
details) we have used square lattices of linear size 
$L\leq 2048$. 
Every simulation starts from a random distribution
of particles.
After a transient regime both models reach a steady state 
characterized by the density of active sites~$\rho_{\scriptscriptstyle \text a}$.
The density~$\rho_{\scriptscriptstyle \text a}$ is the order parameter
and the particle density~$\rho$ is the control parameter
of the absorbing phase transition, i.e., the order parameter
vanishes at the critical density~$\rho_{\text c}$ according to,
$\rho_{\scriptscriptstyle \text a} \propto \delta\rho^{\beta}$ ,
with the reduced control parameter 
$\delta\rho=\rho/\rho_{\text c}-1$. 
Additionally to the order parameter we consider its 
fluctuations $\Delta \rho_{\scriptscriptstyle \text a}$.
Approaching the transition point from above ($\delta\rho>0$) 
the fluctuations diverge according to (see~\cite{LUEB_22,LUEB_24})
$\Delta \rho_{\scriptscriptstyle \text a}\propto  \delta\rho^{-\gamma^{\prime}}$.
Below the critical density (in the absorbing phase)
the order parameter as well as its fluctuations
are zero in the steady state.

It was shown recently that the order parameter as well as
its fluctuations obey the scaling forms~\cite{LUEB_26}
\begin{eqnarray}
\label{eq:scal_ansatz_EqoS}
\rho_{\scriptscriptstyle \text a}(\delta\rho, h) 
\; & \sim & \; 
\lambda^{-\beta}\, \, {\tilde R}
(a_{\scriptscriptstyle \rho}  
\delta \rho \; \lambda, a_{\scriptscriptstyle h} h \;
\lambda^{\sigma}) \, ,\\
\label{eq:scal_ansatz_Fluc}
a_{\scriptscriptstyle \Delta} \,
\Delta \rho_{\scriptscriptstyle \text a}(\delta\rho, h) 
\; & \sim & \; 
\lambda^{\gamma^{\prime}}\, \, {\tilde D}
(a_{\scriptscriptstyle \rho} \delta \rho \; \lambda, 
a_{\scriptscriptstyle h} h \, \lambda^{\sigma})  \, ,
\end{eqnarray}
where $h$ denotes an external field which is conjugated
to the order parameter~\cite{LUEB_22}.
The universal scaling functions ${\tilde R}(x,y)$ and 
${\tilde D}(x,y)$ are the same for all systems belonging 
to a given universality class whereas all non-universal 
system-dependent features (e.g.~the lattice structure, 
the update scheme, etc.)
are contained in the so-called non-universal metric factors 
$a_{\scriptscriptstyle \rho}$, $a_{\scriptscriptstyle h}$,
and $a_{\scriptscriptstyle \Delta}$~\cite{PRIVMAN_1}.
The universal scaling functions are normed
by the conditions ${\tilde R}(1,0)={\tilde R}(0,1)={\tilde D}(0,1)=1$
and the non-universal metric factors can be determined
from the amplitudes of 
\begin{eqnarray}
\label{eq:metric_factors_a_rho}
\rho_{\scriptscriptstyle \text a}(\delta \rho, h=0) \; & \sim & \; 
(a_{\scriptscriptstyle \rho} \, \delta \rho)^{\beta} \, ,\\
\label{eq:metric_factors_a_h}   
\rho_{\scriptscriptstyle \text a}(\delta \rho =0, h) \; & \sim & \; 
(a_{\scriptscriptstyle h} \, h)^{\beta / \sigma} \, , \\
\label{eq:metric_factors_a_Delta}   
a_{\scriptscriptstyle \Delta} \,
\Delta\rho_{\scriptscriptstyle \text a}(\delta \rho=0, h) \; & \sim &\; 
(a_{\scriptscriptstyle h} \, h)^{-\gamma^{\prime}/\sigma} \, .
\end{eqnarray}
These equations are obtained by choosing
in the scaling forms [Eqs.\,(\ref{eq:scal_ansatz_EqoS},
\ref{eq:scal_ansatz_Fluc})] 
$a_{\scriptscriptstyle \rho} \delta\rho \, \lambda=1$ 
and $a_{\scriptscriptstyle h} h \, \lambda^{\sigma}=1$, respectively.

\begin{figure}[t]
  \includegraphics[width=8.0cm,angle=0]{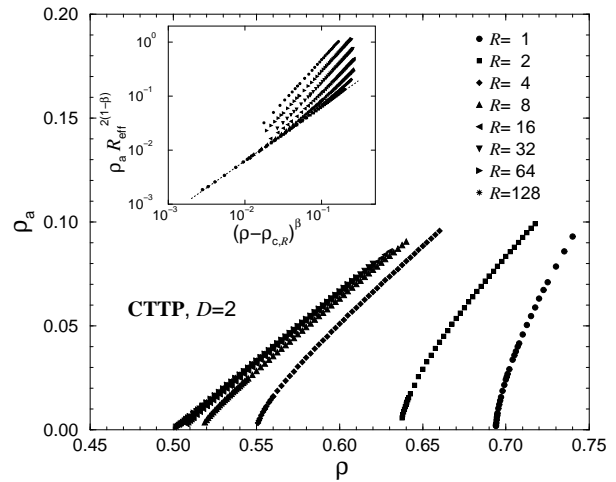}
  \caption{
    The order parameter of the CTTP 
    for various values of the interaction range~$R$.
    With increasing interaction range the critical density
    tends to the mean-field value 
    $\rho_{\scriptscriptstyle {\text c}, R\to \infty}=1/2$.
    The inset displays the order parameter which is
    rescaled according to
    Eq.\,(\protect\ref{eq:rho_a_zf_asymp}).
    The dotted line correspond to the scaling
    laws $y=m\, x$ with 
    $m=(a_{\scriptscriptstyle \rho, R=1}/
    \rho_{\scriptscriptstyle {\text c}, R=1})^{\beta_{\scriptscriptstyle D=2}}$.
    }
  \label{fig:cross_over_01} 
\end{figure}

Usually scaling functions are only known above the
upper critical dimension~$D_{\scriptscriptstyle \text c}$
where the mean-field theory applies.
In the case of the CLG model and CTTP the mean-field
scaling functions are given by~\cite{LUEB_25,LUEB_26}
${\tilde R}_{\scriptscriptstyle \text MF}  
(x, y) = 
{x}/{2} +  (y  + ({x}/{2})^2  )^{1/2} $
as well as
${\tilde D}_{\scriptscriptstyle \text MF}  
(x , y) = {{\tilde R}_{\scriptscriptstyle \text MF}(x,y)}  /  
(y  + ( {x}/{2} )^2 )^{1/2} $
i.e., the mean-field exponents are 
$\beta_{\scriptscriptstyle \text MF} =1$, 
$\sigma_{\scriptscriptstyle \text MF}=2 $, and
$\gamma^{\prime}_{\scriptscriptstyle \text MF} =0$
(corresponding to a finite jump of the fluctuations).
Below the upper critical dimension the universal
scaling functions depend on the dimension and 
are unknown due to a lack of analytical solutions.

   
In the original CLG model and the original
CTTP particles of active sites are moved to nearest
neighbors only, i.e., the range of interactions 
is $R=1$.
In the following we consider a modified CLG model
and a modified CTTP where particles of active sites
are moved (according to the rules of each model) 
to randomly selected sites within a 
radius~$R$. 
The order parameter is plotted 
in Fig.\,\ref{fig:cross_over_01} 
for various ranges of interactions ($R\in \{1,2,4,\ldots,128 \}$).
In the following
we examine how the varying 
interaction range 
affects the scaling behavior in the vicinity
of the absorbing phase transition
which now takes place at the critical 
density~$\rho_{\scriptscriptstyle {\text c}, R}$.


The crossover scaling function at zero field
has to incorporate the range of interactions as an additional
scaling field. We make the phenomenological
ansatz
\begin{equation}
\label{eq:scal_ansatz_EqoS_co}
\rho_{\scriptscriptstyle \text a} 
(\rho, R_{\scriptscriptstyle \text {eff}}) 
\;  \sim \;  
\lambda^{-\beta_{\scriptscriptstyle \text {MF}}} 
\, \, {\tilde {\EuFrak R}}
({\EuFrak a}_{\scriptscriptstyle \rho} (\rho-\rho_{\scriptscriptstyle {\text c}, R}) 
\; \lambda, 
{\EuFrak a}_{\scriptscriptstyle \text R}^{-1} R_{\scriptscriptstyle \text {eff}}^{-1} 
\; \lambda^{\phi} 
) , 
\end{equation}
where the scaling function ${\tilde {\EuFrak R}}$ is universal since we
allow for the non-universal metric factors 
${\EuFrak a}_{\scriptscriptstyle \rho}$
and ${\EuFrak a}_{\scriptscriptstyle \text R}$.
The Ginzburg criterion states that the mean-field picture
is self-consistent in the active phase as long as the
fluctuations within a correlation volume are small
compared to the order parameter itself (see~\cite{ALSNIELSEN_1}).
Thus the crossover exponent is given by
$\phi = 
({2 \beta_{\scriptscriptstyle {\text MF}} -
\nu_{\scriptscriptstyle {\text MF}} D})/{D}
 =  ({4- D})/{2 D}$,
where $\nu_{\scriptscriptstyle {\text MF}}=1/2$
denotes the critical exponents of the spatial correlation
length.
In order to avoid lattice effects we 
use the effective interaction
range~\cite{MON_1}
\begin{equation}
R_{\scriptscriptstyle \text {eff}}^2 \; = \;
\frac{1}{z} \, \sum_{i \neq j}  
| {\underline r}_{\scriptscriptstyle i} \, - \,
{\underline r}_{\scriptscriptstyle j}|^2 \, , \quad
| {\underline r}_{\scriptscriptstyle i} \, - \,
{\underline r}_{\scriptscriptstyle j}| \le R
\label{eq:def_R_eff}
\end{equation}
where $z$ denotes the number of lattice sites within
a radius~$R$ (see Table~\ref{table:radius}).
The mean-field scaling behavior should be recovered
for $R\to \infty$, thus
\begin{equation}
{\tilde {\EuFrak R}}(x,0) \; = \; {\tilde R}_{\scriptscriptstyle {\text MF}}(x,0)
\; = \; x^{\beta_{\scriptscriptstyle {\text MF}}}
\label{eq:mf_limit}
\end{equation}
which implies
${\EuFrak a}_{\scriptscriptstyle \rho} \; = \;
{a_{\scriptscriptstyle \rho, R\to\infty}} /
{\rho_{\scriptscriptstyle {\text c}, R\to\infty}}$
These factors were already determined in previous
works where absorbing phase transitions with infinite 
particle hopping were investigated~\cite{LUEB_25}. 
The non-universal metric factor~${\EuFrak a}_{\scriptscriptstyle \text R}$
has to be determined by a second condition.
Several ways are possible (e.g.~${\tilde {\EuFrak R}}(0,1)=1$)
but for the sake of 
convenience we force that ${\tilde {\EuFrak R}}$ scales as
\begin{equation}
{\tilde {\EuFrak R}}(x,1) \; \sim \;
x^{\beta_{\scriptscriptstyle D}} \, ,
\quad\quad {\text {for}} \quad x\to 0 , 
\label{eq:fractur_R_small_x}
\end{equation}
where $\beta_{\scriptscriptstyle D}$ denotes the 
non-mean-field order parameter exponent of the 
corresponding $D$-dimensional system.
Setting ${\EuFrak a}_{\scriptscriptstyle \text R}^{-1} 
R_{\scriptscriptstyle \text {eff}}^{-1} 
\; \lambda^{\phi} =1$ in Eq.\,(\ref{eq:scal_ansatz_EqoS_co})
yields for zero field
\begin{equation}
\label{eq:scal_plot_EqoS_co}   
\rho_{\scriptscriptstyle \text a} 
(\rho, R_{\scriptscriptstyle \text {eff}}) 
\;  \sim \;  
({\EuFrak a}_{\scriptscriptstyle \text R} 
R_{\scriptscriptstyle \text {eff}})^{-\beta_{\scriptscriptstyle \text {MF}}/\phi} 
\, \, {\tilde {\EuFrak R}}
({\EuFrak a}_{\scriptscriptstyle \rho} (\rho-\rho_{\scriptscriptstyle {\text c}, R}) 
\; 
{\EuFrak a}_{\scriptscriptstyle \text R}^{1/\phi}
R_{\scriptscriptstyle \text {eff}}^{1/\phi} 
,1) 
\end{equation}
Taking into account that the $D$-dimensional scaling
behavior is recovered for $R=1$ we find
\begin{equation}
{\EuFrak a}_{\scriptscriptstyle \text R} \; = \;
\left ( \frac{\rho_{\scriptscriptstyle {\text c},R=1}}
{a_{\scriptscriptstyle \rho, R=1}}\,
\frac{a_{\scriptscriptstyle \rho,R\to \infty}}
{\rho_{\scriptscriptstyle {\text c},R\to \infty}}
\right )^{{\phi \beta_{\scriptscriptstyle D}}/
{(\beta_{\scriptscriptstyle \text MF}-\beta_{\scriptscriptstyle D}})}  .
\label{eq:metric_factors_fractur_R}
\end{equation}

\begin{figure}[t]
  \includegraphics[width=8.0cm,angle=0]{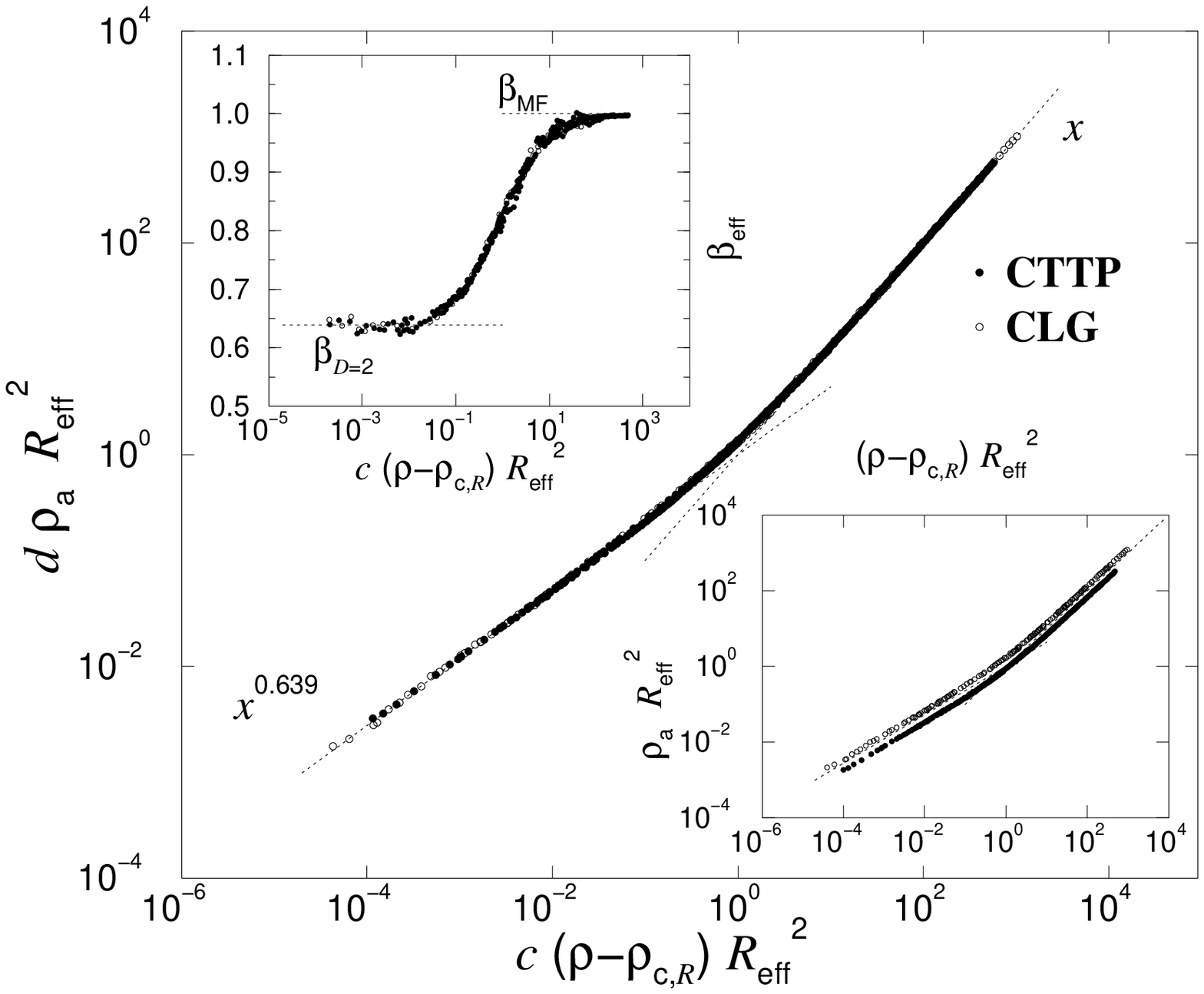}
  \caption{
    The rescaled order parameter.
    The metric factors are given by 
    $c={\EuFrak a}_{\scriptscriptstyle \rho} 
    {\EuFrak a}_{\scriptscriptstyle \text R}^2$ and 
    $d={\EuFrak a}_{\scriptscriptstyle \text R}^2$.
    The data of both models display an excellent 
    collapse to the universal crossover scaling function 
    ${\tilde {\EuFrak R}}(x,1)$.
    The dashed lines correspond to the asymptotic behavior
    of the two-dimensional system 
    ($\beta_{\scriptscriptstyle D=2}=0.639$~\protect\cite{LUEB_24})
    and of the mean-field behavior
    ($\beta_{\scriptscriptstyle \text {MF}}=1$).
    Neglecting the metric factors each model is characterized
    by its own scaling function (see lower left inset).
    The upper right inset displays the effective exponents $\beta_{\scriptscriptstyle \text
    {eff}}$ for both considered models.
    The data of the CLG model and of the CTTP
    exhibit an excellent data collapse.
   }
  \label{fig:uni_co_opzf} 
\end{figure}

According to the above scaling form we plot in 
Fig.\,\ref{fig:uni_co_opzf} the rescaled order parameter 
$\rho_a\,({\EuFrak a}_{\scriptscriptstyle \text R} 
R_{\scriptscriptstyle \text {eff}})^{2}$ as a function of 
the rescaled control parameter 
${\EuFrak a}_{\scriptscriptstyle \rho} (\rho-\rho_{\scriptscriptstyle {\text c}, R})
({\EuFrak a}_{\scriptscriptstyle \text R} 
R_{\scriptscriptstyle \text {eff}})^{2}$
for the two-dimensional ($\phi=1/2$) CLG model and the 
two-dimensional CTTP.
The values of the metric factors are listed in 
Table\,\ref{table:critical_indicees} and are determined from data of 
previous simulations (via a direct measurement of the amplitudes of the
corresponding power-laws).
Thus no parameter fitting is applied.
We observe an excellent data collapse for the entire
range of the crossover confirming the phenomenological
ansatz.
In the inset of Fig.\,\ref{fig:uni_co_opzf} we plot
the same data without metric factors.
As can be seen each model is characterized by its
own scaling function.

\begin{table}[t]
\caption{The range of interactions $R$, the corresponding
number of next neighbors~$z$ on a square lattice and 
the effective range of 
interactions~$R_{\scriptscriptstyle \text {eff}}$ for which 
we have carried out simulations.
Additionally, the values of the critical densities are listed.}
\label{table:radius}
\begin{tabular}{lllll}
$R$ & $z_{\scriptscriptstyle D=2}$ 
& $R_{\scriptscriptstyle \text {eff}, D=2}^2$
& $\rho_{\scriptscriptstyle {\text c}, R}^{\scriptscriptstyle \text {CLG}}\quad\quad\quad\quad$ 
& $\rho_{\scriptscriptstyle {\text c}, R}^{\scriptscriptstyle \text {CTTP}}\quad$\\
\colrule 
1 	   & 4 	    & 1                         & 0.34494(3)     & 0.69392(1)   \\
2 	   & 12     & $\frac{7}{3}$             & 0.22432(4)     & 0.63649(2)   \\
4 	   & 48     & 8                         & 0.16802(7)     & 0.55005(3)   \\
8 	   & 196    & $\frac{1546}{49}$         & 0.14050(9)     & 0.51688(4)   \\
16 	   & 796    & $\frac{25274}{199}$       & 0.12977(10)     & 0.50552(6)   \\
32 	   & 3208   & $\frac{204875}{401}$      & 0.12598(11)     & 0.50161(7)   \\
64  	   & 12852  & $\frac{13146247}{6426}$   & 0.12499(16)     & 0.50046(8)   \\
128$\quad\quad$ & 51432\quad\quad  & $\frac{105255421}{12858}\quad\quad$ & 0.12465(19) & 0.50019(9)\\
\end{tabular}
\end{table}

Since the entire crossover region covered several 
decades it could be difficult to observe small but systematic
differences between the scaling functions of both models.
It its therefore instructive to examine the crossover
via the so-called effective exponent~\cite{LUIJTEN_2}
\begin{equation}
\beta_{\scriptscriptstyle \text {eff}} \; = \;
\frac{\partial\hphantom{\ln{x}}}{\partial \ln{x}}
\, \ln{{\tilde{\EuFrak R}}(x,1)} .
\label{eq:def_eff_exp_beta}
\end{equation}
The corresponding data are shown in Fig.\,\ref{fig:uni_co_opzf}.
The excellent data collapse 
of~$\beta_{\scriptscriptstyle \text {eff}}$ of both models 
over more than 7 decades strongly supports the hypothesis
that the crossover function is a universal function.

\begin{figure}[t]
  \includegraphics[width=8.0cm,angle=0]{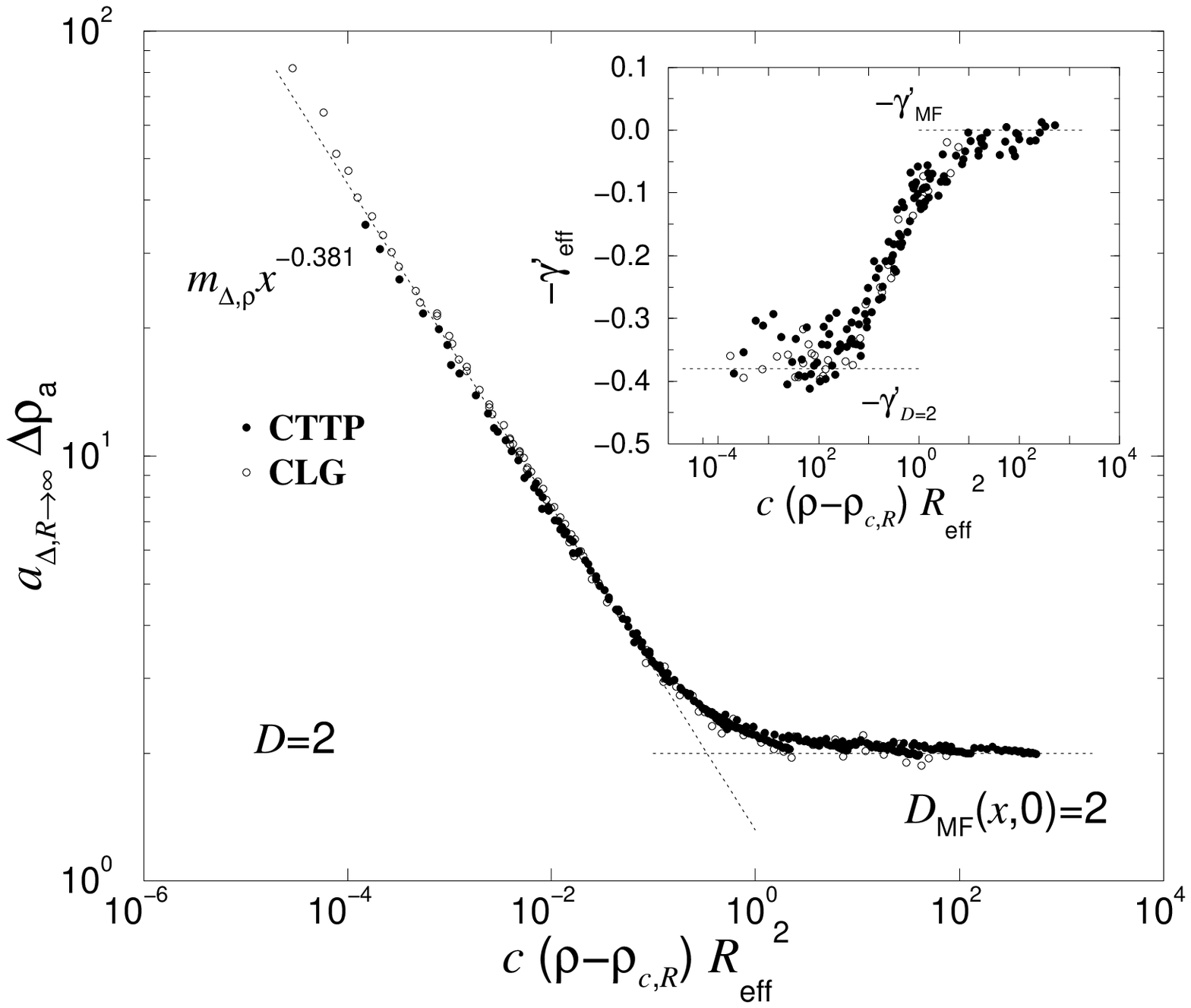}
  \caption{
    The rescaled fluctuations of the order parameter.
    The metric factor is given by 
    $c={\EuFrak a}_{\scriptscriptstyle \rho} 
    {\EuFrak a}_{\scriptscriptstyle \text R}^2$.
    The data of both models display a good 
    collapse to the universal crossover scaling function 
    ${\tilde {\EuFrak D}}(x,1)$.
    The dashed lines correspond to the asymptotic behavior
    of the two-dimensional system 
    ($\gamma^{\prime}_{\scriptscriptstyle D=2}=0.381$~\protect\cite{LUEB_24}
    and $m_{\scriptscriptstyle \Delta,\rho}=1.28$)
    and of the mean-field behavior
    ($\gamma^{\prime}_{\scriptscriptstyle \text {MF}}=0$).
    The inset displays the corresponding effective
    exponent~$\gamma^{\prime}_{\scriptscriptstyle \text {eff}}$.
   }
  \label{fig:uni_co_flzf} 
\end{figure}

We now consider the order parameter fluctuations.
Analogous to the order parameter we make the scaling
ansatz ($\gamma^{\prime}_{\scriptscriptstyle {\text MF}}=0$)
\begin{equation}
\label{eq:scal_ansatz_Fluc_co}
{\EuFrak a}_{\scriptscriptstyle \Delta} \,
\Delta\rho_{\scriptscriptstyle \text a}
(\rho, R_{\scriptscriptstyle \text {eff}}) 
\; \sim \; 
{\tilde {\EuFrak D}}
({\EuFrak a}_{\scriptscriptstyle \rho} (\rho-\rho_{\scriptscriptstyle {\text c}, R})
\; \lambda, 
{\EuFrak a}_{\scriptscriptstyle \text R}^{-1} R_{\scriptscriptstyle \text {eff}}^{-1} 
\; \lambda^{\phi} 
) . 
\end{equation}
Again the mean-field behavior should
be recovered for $R\to\infty$, implying
${\tilde {\EuFrak D}}(x,0)  =  
{\tilde D}_{\scriptscriptstyle {\text MF}}(x,0)=2\, $
as well as 
${\EuFrak a}_{\scriptscriptstyle \Delta} = 
{a_{\scriptscriptstyle \Delta, R\to\infty}}$.
Setting ${\EuFrak a}_{\scriptscriptstyle \text R}^{-1} 
R_{\scriptscriptstyle \text{eff}}^{-1} \; 
\lambda^{\phi} =1$ yields 
\begin{equation}
{\EuFrak a}_{\scriptscriptstyle \Delta} \,
\Delta\rho_{\scriptscriptstyle \text a}
(\rho, R_{\scriptscriptstyle \text {eff}}) 
\; \sim \; 
{\tilde {\EuFrak D}} ({\EuFrak a}_{\scriptscriptstyle \rho} (\rho-\rho_{\scriptscriptstyle
{\text c}, R}) \; {\EuFrak a}_{\scriptscriptstyle \text R}^{1/\phi} 
R_{\scriptscriptstyle\text {eff}}^{1/\phi},1) . 
\label{eq:scal_plot_Fluc_co}
\end{equation}
For finite $R$, the fluctuations diverge at the critical point, i.e.,
the universal function ${\tilde {\EuFrak D}}$ scales as
\begin{equation}
{\tilde {\EuFrak D}}(x,1) \; \sim \; m_{\scriptscriptstyle\Delta,\rho} \,
x^{-\gamma^{\prime}_{\scriptscriptstyle D}} \, ,
\quad\quad {\text {for}} \quad x\to 0 \, . 
\label{eq:fractur_D_small_x}
\end{equation}
The universal amplitude~$m_{\scriptscriptstyle\Delta,\rho}$
can be determined in the following way:~The scaling form
Eq.\,(\ref{eq:scal_ansatz_Fluc_co}) has to equal for $R=1$
the $D$-dimensional scaling behavior 
[see Eq.\,(\ref{eq:scal_ansatz_Fluc})]
\begin{equation}
\Delta \rho_{\scriptscriptstyle\text {a}} \; \sim \;
a_{\scriptscriptstyle \Delta,R=1}^{-1} \; {\tilde D}(1,0) \;
\left ( a_{\scriptscriptstyle \rho,R=1} \,
\frac{(\rho-\rho_{\scriptscriptstyle
{\text c}, R=1})}
{\rho_{\scriptscriptstyle {\text c}, R=1}}
\right )^{-\gamma^{\prime}_{D}} \, .
\label{eq:scaling_fluc_R1}
\end{equation}
Thus we find 
\begin{equation}
\label{eq:uni_ampl_m_delta}   
m_{\scriptscriptstyle \Delta,\rho}  = 
{\tilde D}(1,0) \,
\frac{a_{\scriptscriptstyle \Delta, R\to \infty}}
{a_{\scriptscriptstyle \Delta, R=1}} \;
\left ( 
\frac{\rho_{\scriptscriptstyle {\text c},R=1}}
{a_{\scriptscriptstyle \rho, R=1}}\,
\frac{a_{\scriptscriptstyle \rho,R\to \infty}}
{\rho_{\scriptscriptstyle {\text c},R\to \infty}}
\right )^\frac{{\gamma^{\prime}_{\scriptscriptstyle D} 
\beta_{\scriptscriptstyle {\text MF}}}}
{{\beta_{\scriptscriptstyle \text MF}-\beta_{\scriptscriptstyle D}}}  
\end{equation}
where the value of the universal scaling function 
${\tilde D}(1,0)=1.87\pm 0.11$ is obtained via direct measurements
of the corresponding two-dimensional systems. 
According to the scaling form Eq.\,(\ref{eq:scal_plot_Fluc_co})
we plot in Fig.\,\ref{fig:uni_co_flzf} the rescaled
fluctuations as a function of the rescaled control
parameter for the two-dimensional CLG model as well as
for the CTTP.
We observe again a good collapse of the data over 
the entire region of the crossover.
Furthermore, both asymptotic behaviors are recovered,
confirming the scaling ansatz Eq.\,(\ref{eq:scal_ansatz_Fluc_co}).

The corresponding effective exponent
$\gamma^{\prime}_{\scriptscriptstyle \text {eff}} = 
{\partial}\ln{{\tilde{\EuFrak D}}(x,1)}
/{\partial \ln{x}}$
is displayed in the inset of Fig.\,\ref{fig:uni_co_flzf}.
Although the data of the effective exponent 
are suffering from statistical fluctuations
one can see that both models are characterized by the 
same universal behavior.

At the end we consider the critical amplitudes
of the scaling functions.
Using the above discussed scaling forms it easy to
show that these amplitudes display  a singular dependence 
on the range of interactions.
For instance the order parameter scales 
sufficiently close to the transition point ($x\to 0$) as
[see Eqs.\,(\ref{eq:fractur_R_small_x},
\ref{eq:scal_plot_EqoS_co},\ref{eq:metric_factors_fractur_R})]
\begin{equation}
\label{eq:rho_a_zf_asymp}
\rho_{\scriptscriptstyle \text a} 
(\rho, R_{\scriptscriptstyle \text {eff}}) 
\; \sim  \;
R_{\scriptscriptstyle \text {eff}}^{(\beta_{\scriptscriptstyle D}
-\beta_{\scriptscriptstyle \text {MF}})/\phi} \,
\left ( a_{\scriptscriptstyle \rho,R=1} 
\frac{\, \rho-\rho_{\scriptscriptstyle {\text c}, R} \, }
{\rho_{\scriptscriptstyle {\text c},R=1}} 
\right )^{\beta_{\scriptscriptstyle D}}  \, .
\end{equation}
Thus this scaling law and the corresponding scaling law for the
fluctuations are only valid for 
finite interaction ranges whereas they become useless
for infinite $R$, signaling the change in the
universality class for $R\to \infty$.
This amplitude scaling can be observed in simulations.
The inset of Fig.\,\ref{fig:cross_over_01} shows the
corresponding data for the CTTP. 
As can be seen, the data of various interaction ranges 
tend to the same power-law behavior if one approaches
the transition point.

\begin{table}[b]
\caption{The non-universal metric factors determined
from previous simulations
via direct measurements of the corresponding power-laws.}
\label{table:critical_indicees}
\begin{tabular}{lcccccc}
 & $\rho_{\scriptscriptstyle {\text c}, R=1}\;$ 
 & $a_{\scriptscriptstyle \rho, R=1}\;$
 & $a_{\scriptscriptstyle \Delta, R=1}\;$
 & $\rho_{\scriptscriptstyle {\text c}, R\to \infty}\;$
 & $a_{\scriptscriptstyle \rho, R\to \infty}\;$
 & $a_{\scriptscriptstyle \Delta, R\to \infty}\;$ \\
\colrule 
${\text {CLG}}_{\scriptscriptstyle  D=2}$ 
& 0.34494 & 0.5089 & 15.50 & 0.1244 & 0.1635	&12.02	\\
${\text {CTTP}}_{\scriptscriptstyle D=2}$
& 0.69392 & 0.3410 & 50.18 & 1/2 & 0.3345 & 24.85 \\
\end{tabular}
\end{table}


In conclusion, the crossover from mean-field to non-mean-field
scaling behavior is numerically investigated for two 
different models exhibiting an absorbing phase transition.
Increasing the range of interactions we are able 
to cover the full crossover
region which spans several decades of the control
parameter.
The excellent collapse of the effective exponents of 
both models strongly supports the interpretation
of the crossover scaling functions in terms of universality,
i.e., the crossover function is universal.

I would like to thank
P.\,K.~Mohanty for useful comments.
This work was financially supported by the 
Minerva Foundation (Max Planck Gesellschaft).

\end{document}